\documentclass[aip,jcp,preprint]{revtex4-1}
\usepackage{graphicx}
\usepackage[svgnames]{xcolor}
\usepackage[colorlinks=true,citecolor=blue,linkcolor=blue,citebordercolor=white,linkbordercolor=white]{hyperref}

\def\eq#1{{Eq.~(\ref{#1})}}

\def\fig#1{{Fig.~\ref{#1}}}

\begin{document}

\author{S.V. Novikov}
\affiliation{A.N. Frumkin Institute of
Physical Chemistry and Electrochemistry, Leninsky prosp. 31,
Moscow 119071, Russia}
\affiliation{National Research University Higher School of Economics, Myasnitskaya Ulitsa 20, Moscow 101000, Russia}

\title[Two-dimensional bimolecular recombination]{Two-dimensional bimolecular recombination in amorphous  organic semiconductors}

\begin{abstract}
We consider the two-dimensional bimolecular recombination of charge carriers in amorphous organic semiconductors having the lamellar structure. We calculate the dependence of the effective recombination rate constant on the carrier density taking into account the correlated nature of the energetic disorder typical for organic semiconductors. Resulting recombination kinetics demonstrates a very rich variety of behaviors depending on the correlation properties of the particular semiconductor and relevant charge density range.
\end{abstract}

\maketitle

\section{Introduction}
Charge carrier recombination is an important process to a very large extent defining the properties and efficacy of semiconductor electronic and optoelectronic devices.\cite{grundmann:book,brutting:book,Fabregat-Santiago:9083} In amorphous organic semiconductors with slow carrier motion the limiting stage in the non-geminate recombination is usually the transport of carriers of the opposite signs to each other, and the act of recombination could be considered as the diffusion-limited bimolecular reaction
\begin{equation}
{\rm h}^+ +{\rm e}^- \rightarrow {\rm neutral},
\label{recomb-reaction}
\end{equation}
(Langevin recombination). For the spatially uniform carrier distribution  reaction (\ref{recomb-reaction}) is described by the second-order kinetic equation
\begin{equation}\label{kin}
    \frac{dn}{dt}=-\gamma n^2.
\end{equation}
We assume here that the densities of electrons and holes are equal $n=p$ and the intrinsic carrier density is negligible.

Rate constants for the diffusion-limited reaction $\rm A +\rm B\rightarrow \rm C$ may be calculated in the space of arbitrary dimension $d$ using the Smoluchowski-Debye method.\cite{Rice:book} In this method the rate constant for the reaction between two particles interacting by the potential energy $U(\vec{r})$ is calculated from the stationary solution of the $d$-dimensional diffusion equation
\begin{equation}\label{eq_SD}
\frac{\partial \rho}{\partial t}=D\frac{\partial}{\partial \vec{r}}\left(\frac{\partial\rho}{\partial \vec{r}}+\beta\frac{\partial U}{\partial \vec{r}}\rho\right),
\end{equation}
where $D=D_A+D_B$ is the sum of the diffusivities of the reacting particles and $\beta=1/kT$; later we consider exclusively the case of the spherically symmetric potential $U(r)$. For the function $\rho_s(r)=\rho(r,t\rightarrow\infty)$ we have the boundary condition
\begin{equation}\label{k_gem}
\gamma=k_g\rho_s(R)=S_d D R^{d-1}\left.\left[\frac{\partial \rho_s}{\partial r}+\beta\rho_s\frac{\partial U}{\partial r}\right]\right|_{r=R},
\end{equation}
where $R$ is the radius of the sphere where the reaction takes place, $S_d$ is the surface of the $d$-dimensional unit sphere, and $k_g$ is the rate constant of the short distance quasi-geminate recombination. In this paper we consider the simplest case of the instant geminate recombination $k_g\rightarrow\infty$ and this condition reduces to $\rho_s(R)=0$. We set also another boundary condition  $\rho_s(R_l)=1$ at some large distance $R_l$. Function $\rho_s(r)=n_B(r)/n_B(R_l)$ may be considered as the ratio of the densities of the particles B if we fix the particle A at the origin, and the boundary at $r=R_l$ serves as the inexhaustible supply of particles B. Spherically symmetric stationary solution of \eq{eq_SD} could be easily found and the formal expression for the rate constant in arbitrary dimension $d$ is
\begin{equation}\label{gamma_d}
\gamma[U]=S_d D \frac{\exp\left[\beta U(R_l)\right]}{\lambda \exp\left[\beta U(R)\right]+\int^{R_l}_R\frac{dr}{r^{d-1}}\exp\left[\beta U(r)\right]},
\end{equation}
where $\lambda=S_d D/k_g$ and later we consider only the case $\lambda=0$ (instant quasi-geminate recombination).

For the traditional Smoluchowski-Debye method we finally let  $R_l\rightarrow\infty$, thus considering the rate constant for the infinite dilution $n\rightarrow 0$. It is well known that for the low dimensionality $d < 3$ this approach fails: integral in the denominator of \eq{gamma_d} diverges and the rate constant goes to zero.\cite{Emeis:2246,Freeman:6002}

The physical reason for the breakdown of the Smoluchowski-Debye method could be understood from the simple argument. The stationary solution could exist if the outflow of the substance by the reaction $\propto \gamma t$ for $t\rightarrow\infty$  is negligible in comparison to the possible diffusional supply  proportional to the accessible $d$-dimensional volume $\propto (D t)^{d/2}$, and the only possibility to fulfill this condition is $d/2 > 1$ and $d > 2$. Diffusion in the low dimensional space cannot provide enough substance for the sustenance of the true stationary rate constant.

This consideration suggests the remedy for $d < 3$: we should set the boundary condition $\rho=1$ not at the infinity but at the distance about the average inter-particle distance which may be estimated as $V_d R_l^dn=1$, where $V_d$ is a volume of the  $d$-dimensional unit ball. Obviously, this procedure introduce a density dependence for $\gamma$ and the recombination is no more the true bimolecular reaction.

As we already noted, charge carrier recombination is of utmost importance for most organic electronic or optoelectronic devices. Yet even in mescopically homogeneous amorphous semiconductors general properties of recombination are still not understood very well. For example, only recently the effect of ubiquitous spatial correlations of the random energy landscape has been considered.\cite{Novikov:22856,Novikov:18854} In organic photovoltaic (OPV) applications the recombination should be prevented by all means in order to get the highest possible efficiency. For this reason a lot of materials have been tested which develop a huge set of various morphologies. A particular type of the mesoscopic arrangement is the formation of the lamellar structures, quite common in regio-regular oligo- and polythiophene materials.\cite{Servet:1809,Hamidi-Sakr:408,Rivnay:121306} In lamellas there is a significant difference of the charge transport properties in the lamella plane and out of plane: in-plane mobility was found to be hundred times greater than the transversal mobility.\cite{Sirringhaus:685} Hence, with proper accuracy we may try to describe the recombination in lamellar materials as a two-dimensional process.

Such attempts already have been done, but in all studies either the spatially correlated nature of the energetic disorder in organic semiconductors was not taken into account\cite{Greenham:245301,Li:702} or the disorder effects were not explicitly considered at all.\cite{Juska:13303,Nenashev:213304} Correlated energetic disorder is typical for amorphous organic semiconductors and to a very large extent determines bulk transport properties of the mesoscopically homogeneous amorphous materials. We should expect some degree of spatial correlation in lamellas either for the energetic disorder induced in lamellas by the surrounding more amorphous material, analogous to the disorder in the thin transport layer of organic field effect transistors,\cite{Veres:199,Veres:4543,Fishchuk:18} or for the intrinsic disorder originated from the irregularities of the lamella structure.   For the two-dimensional recombination in the case of the spatially uncorrelated disorder the rate constant essentially follows the carrier mobility,\cite{Greenham:245301} exactly as in the case of three-dimensional recombination where the Langevin formula for the recombination rate constant still holds
\begin{equation}\label{Lan}
\gamma_{\rm L}=\frac{4\pi e}{\varepsilon}\left(\mu_{+}+\mu_{-}\right),
\end{equation}
if we consider the actual mobilities $\mu_{+}$, $\mu_{-}$ of holes and electrons in the amorphous semiconductor; here $\varepsilon$ is the dielectric constant of the medium. If we take into account the spatial correlation of the random energy landscape, the situation becomes different and the bimolecular recombination rate constant $\gamma$ is no more simply proportional to the carrier mobility with some universal proportionality coefficient but depends on other properties of the landscape.\cite{Novikov:22856,Novikov:18854}

Conception of the two-dimension recombination is usually invoked for the explanation of the experimentally observed reduced recombination rate with the reduction factor $\zeta=\gamma/\gamma_{\rm L} \ll 1$ and slow charge density decay kinetics $n(t)$, related to the non-trivial density dependence of $\gamma$.\cite{Juska:13303} The change of the density dependence by the effect of spatial correlation is quite possible. For these reasons the study of the two-dimensional recombination in the spatially correlated random energy landscape is highly desirable.

We are going to show that the approach suggested previously\cite{Novikov:22856,Novikov:18854} can be easily extended to the study of two-dimensional recombination as well (in future we limit our consideration to $d=2$, where $S_2=2\pi$ and $V_2=\pi$). Our approach permits in a very simple way reproduce and extend well known results for the two-dimensional recombination in the ordered medium,\cite{Greenham:245301,Juska:13303,Nenashev:213304} yet the major benefit is the possibility to provide the analytic treatment of the recombination in the disordered materials. Possible application to the recombination in the thiophene-based materials may serve as a good example of this approach, though any bimolecular diffusion-limited bimolecular reaction in 2D geometry in the presence of the energetic disorder with the Gaussian density of states could be considered in the similar way.

\section{Effects of correlated energetic disorder}

As a starting point we can compare our result for $\gamma$ in the absence of the disorder with the previous calculations carried out by the different method.\cite{Greenham:245301,Juska:13303,Nenashev:213304} For the recombination taking into account only the Coulomb interaction between carriers $U(r)=-e^2/\varepsilon r$, and for $n \ll 1/R_{\rm Ons}^2$ (here $R_{\rm Ons}=e^2/\varepsilon kT$ is the Onsager radius)  the integral in \eq{gamma_d} is approximately equal to $\ln (R_l/R_{\rm Ons})= -\frac{1}{2}\ln (\pi R_{\rm Ons}^2 n)$, so the leading logarithmic asymptotics for $\gamma$ is
\begin{equation}\label{low-n}
\gamma\simeq 4\pi D\left(\ln\frac{1}{R_{\rm Ons}^2 n}\right)^{-1},
\end{equation}
exactly the same as obtained previously,\cite{Nenashev:213304} and in the opposite limit the integral can be approximately calculated by the Laplace method giving
\begin{equation}\label{high-n}
\gamma\simeq 2\pi^{3/2} D \left(R_{\rm Ons}^2 n\right)^{1/2},
\end{equation}
which differs from the known results\cite{Greenham:245301,Juska:13303,Nenashev:213304} by numerical coefficients $\simeq O(1)$, reflecting the different treatment of the boundary condition at $r=R_l$. Obviously, this difference is not relevant due to the approximate nature of all approaches, both used previously\cite{Greenham:245301,Juska:13303,Nenashev:213304} and in our paper: in all cases we replace the random mixture of charges by the regular two-dimensional lattice of carriers and assume that the shape of the lattice cell may be approximated by the perfect circle.

The real power of our approach is the ability to incorporate the effects of the energetic disorder. We recently demonstrated how to calculate the three-dimensional recombination rate constant in amorphous organic semiconductors having the Gaussian density of states and assuming that the process of energetic relaxation of carriers is over.\cite{Novikov:22856,Novikov:18854} For simplicity we may assume that mobilities of the carriers of the opposite signs are very different and the slow carriers are sitting in potential wells while fast carriers of the opposite sign are approaching. Then, in addition to the Coulomb attraction, the energies of electron and hole also have random contributions. The crucial approximation giving the possibility to provide a closed expression for the rate constant is the assumption that the random energy landscape $U_{\rm rand}(\vec{r})$ around the carrier sitting at the bottom of the potential well with the depth $U_0$ (we assume that $U_{\rm rand}(0)=-U_0$) may be replaced with the conditionally averaged potential $\left<U_{\rm rand}(\vec{r})\right>=-U_0 C(\vec{r})$ where  $C(\vec{r})$ is the disorder correlation function normalized in such a way that $C(0)=1$. This average is exact for the Gaussian random landscape. The resulting rate constant is then obtained by the averaging of the rate constant for the particular $U_0$ with the density of occupied states. Our approach immediately tells us that the deviation of the 3D recombination rate constant from the Langevin value $\gamma_{\rm L}$ in mesoscopically spatially homogeneous amorphous semiconductors stems not from the energetic disorder per se but from the spatial correlation of the random energy landscape. In agreement, simulation of $\gamma$ for the uncorrelated disorder gives the Langevin rate constant.\cite{Albrecht:455,Groves:155205,Holst:235202}

We may immediately extend the results to the case of the two-dimensional recombination using the rate constant (\ref{gamma_d}), where the potential energy now includes the random contribution (see details in ref. \citenum{Novikov:22856,Novikov:18854}). The total bimolecular recombination rate constant averaged over disorder is
\begin{eqnarray}\label{gamma0}
\gamma = \frac{1}{\left(2\pi\sigma^2\right)^{1/2}} &
\int\limits_{-\infty}^\infty & dU_0 \gamma[U_{\rm eff}] \frac{\exp\left(-\frac{U_0^2}{2\sigma^2}\right)}{1+\exp\left(-\frac{U_0+\mu}{kT}\right)},\\
\label{gamma0-1}
U_{\rm eff}(r) & = & -\frac{e^2}{\varepsilon r}\pm U_0 C(r),
\end{eqnarray}
where $\sigma\simeq 0.1$ eV is the rms disorder. Chemical potential $\mu$ is calculated from the relation
\begin{equation}\label{mu}
\frac{n_R}{\left(2\pi\sigma^2\right)^{1/2}}\int\limits_{-\infty}^\infty dU \frac{\exp\left(-\frac{U^2}{2\sigma^2}\right)}{1+\exp\left(\frac{U-\mu}{kT}\right)}=n
\end{equation}
where the limiting density $n_R$ is equal to $1/\pi R^2$; the signs of $U_0$ and $U$ in the Fermi-Dirac factors in eqs~(\ref{gamma0-1}) and (\ref{mu}) are opposite because the corresponding energy in \eq{gamma0-1} is $-U_0$.

The actual sign before the term $\propto U_0$ in $U_{\rm eff}(r)$ depends on the particular kind of the energetic disorder. The plus sign means that the random energies of electrons and holes move in the opposite directions with the variation of some disorder-governing parameter (the anti-parallel disorder, the example is the electrostatic disorder generated by the randomly located and oriented dipoles and quadrupoles\cite{Novikov:22856}). The minus sign means that the random energies of electrons and holes move in the same direction (the parallel disorder); such landscape could originate from the dominating contribution of the conformational disorder.\cite{Masse:115204,Novikov:18854}

Relaxed slow carriers are concentrated at $U_0 >0$, so the anti-parallel disorder provides effective repulsion between carriers of the opposite signs, acting in the opposite direction to the Coulomb attraction. For very deep wells the total effective interaction between carriers of the opposite signs could even become repulsive. This makes the recombination less efficient and diminishes $\gamma$, while the parallel disorder gives additional effective attraction enhancing the recombination. This means also that for the anti-parallel disorder, providing the additional energetic barrier for recombination, we should expect more drastic effect on the temperature dependence of $\gamma$, while for the parallel disorder the temperature dependence should be mostly determined by the temperature dependence of the diffusivity.

A natural limitation of our approach is the assumption of full relaxation; thus, the recombination of carriers at the short time after excitation is beyond possibilities of our approximation. This means that the presented theory could be applied to rather thick layers in the submicron and micron range and for not very low temperature, where the relaxation process is essentially over,\cite{Devizis:27404,Mozer:35217,Melianas:1806004} or to the steady state conditions.

For the anti-parallel electrostatic disorder we may expect the correlation function $C(r)\propto a/r$, $r\gg a$ (here $a\approx R$ is the intermolecular distance) analogous to  polar amorphous organic materials,\cite{Novikov:2584} while for the parallel conformational disorder the exponential correlation function $C(r)\propto \exp(-r/l)$ is expected; it is observed in many amorphous materials and some organic glasses.\cite{Masse:115204,Novikov:18854}

\begin{figure}[tbp]
\centering
\includegraphics[width=3in]{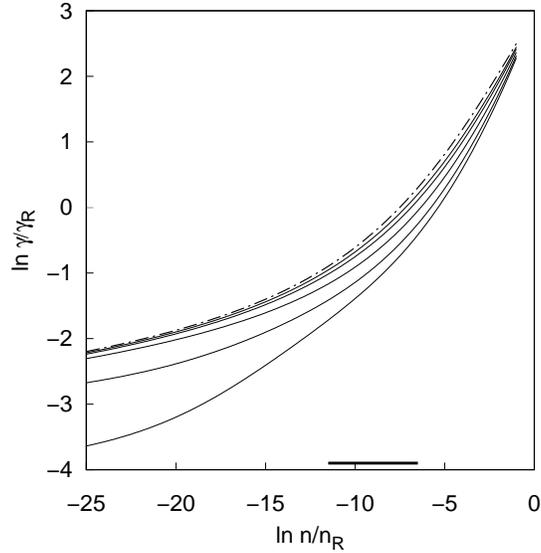}
\caption{Density dependence of $\gamma$ for the dipolar correlation function and  $\sigma$ equals to 0.05 eV, 0.07 eV, 0.1 eV, 0.13 eV, and 0.15 eV, from the top curve downward. Dot-dashed line in this and other figures shows the corresponding dependence for the no disorder case $\sigma=0$. For all curves $\varepsilon=3$, $R=1$ nm, $T=300$ K, and these parameters are used in all figures;  $\gamma_R = 2\pi D$. Thick bar at the abscissa axis in all figures indicates the typical range of $n$ in experiments.}
\label{2d-dip}
\end{figure}

\begin{figure}[tbp]
\centering
\includegraphics[width=3in]{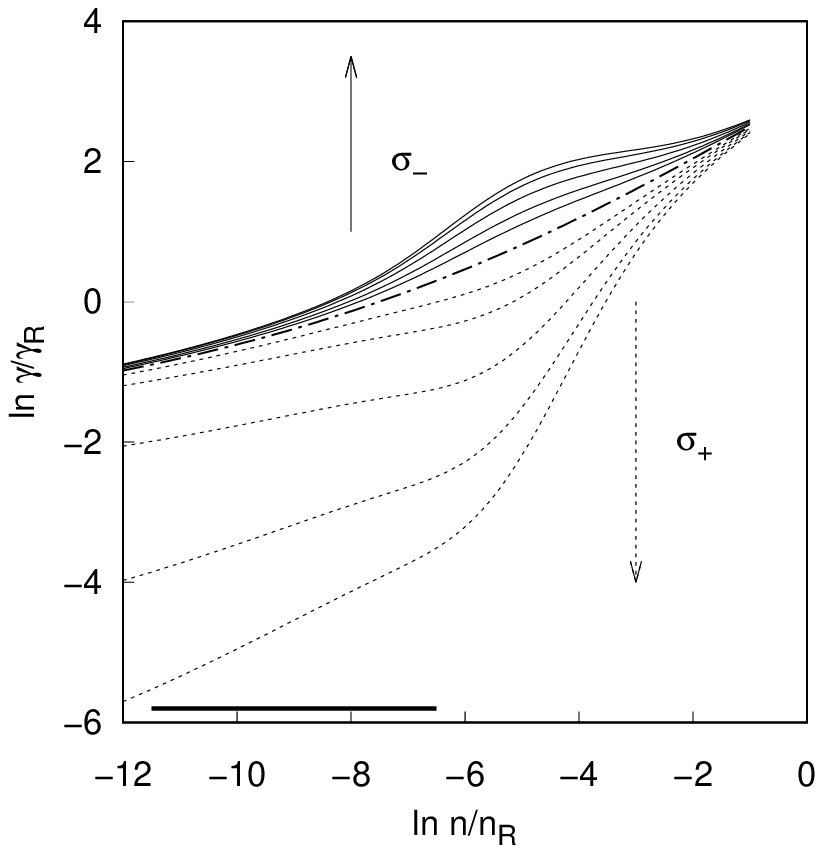}
\caption{Density dependence of $\gamma$ for the exponential correlation function with the correlation length $l=5$ nm and $\sigma$ equals to 0.05 eV, 0.07 eV, 0.1 eV, 0.13 eV, and 0.15 eV. Solid lines show the dependences for the parallel disorder with the minus sign in $U_{\rm eff}(r)$ in \eq{gamma0-1}, while the dashed lines show the corresponding dependences for the anti-parallel disorder with the plus sign in $U_{\rm eff}(r)$. }
\label{2d-exp}
\end{figure}

\begin{figure}[tbp]
\centering
\includegraphics[width=3in]{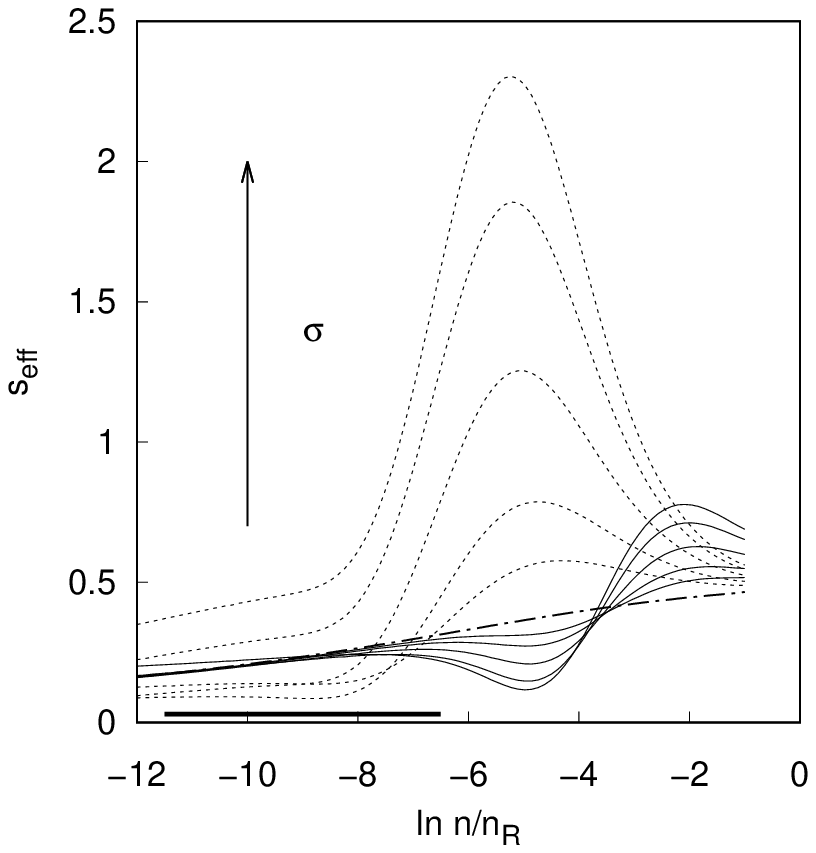}

a)

\includegraphics[width=3in]{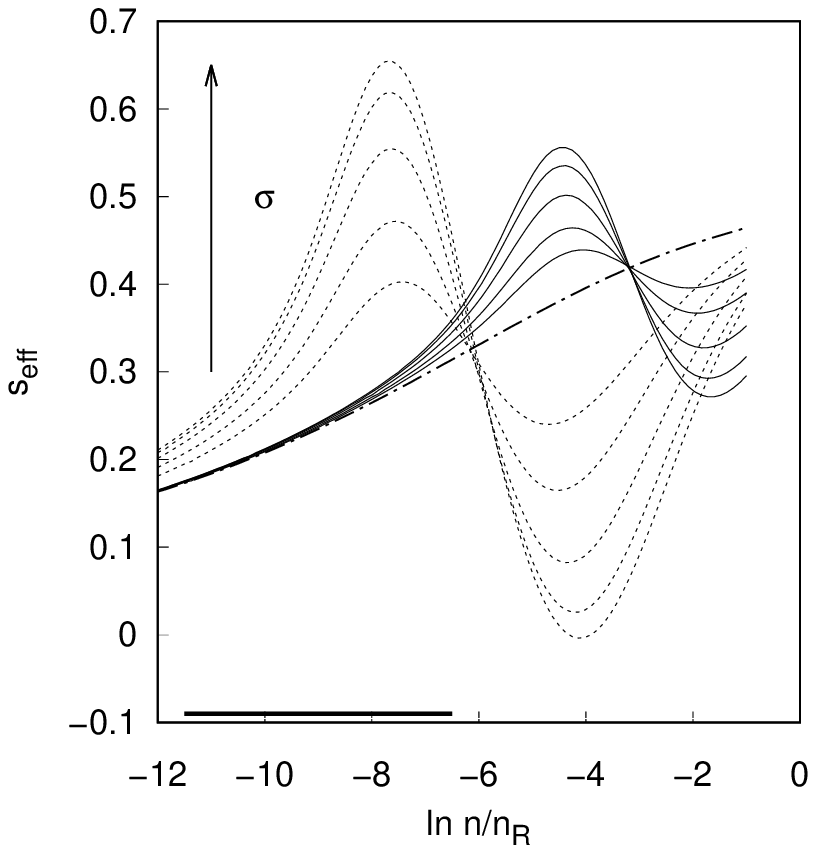}

b)

\caption{Effective exponent $s_{\rm eff}=\frac{d\ln \gamma}{d\ln n}$ for the exponential correlation function and $\sigma$ equals to 0.05 eV, 0.07 eV, 0.1 eV, 0.13 eV, and 0.15 eV, from the bottom curve upwards. Solid lines show $s_{\rm eff}$ for $l=2$ nm and dashed lines for $l=10$ nm, correspondingly. Figure a) shows the dependence for the anti-parallel disorder with the plus sign in $U_{\rm eff}(r)$  in \eq{gamma0-1}, and figure b) shows the corresponding dependence for the parallel disorder with the minus sign.}
\label{s}
\end{figure}

\fig{2d-dip} shows the density dependence of $\gamma$ for the anti-parallel dipolar disorder having the correlation function $C(r)= A a/r$, $r\gg a$ with $A\simeq 1$ (for example, for the particular model of the simple cubic lattice with sites occupied by dipoles $A=0.76..$\cite{Dunlap:80}). \fig{2d-exp} shows the corresponding dependence for the exponential correlation function $C(r)=\exp(-r/l)$, and for locally ordered lamellas $l=2 - 5$ nm is plausible. Both figures show that in the limited carrier density range the density dependence of $\gamma$ may be approximated as $\gamma\propto n^{s}$ (the typical variation of $n(t)$ observed in experiments is about one order of magnitude\cite{Sliauzys:224,Pivrikas:677}). \fig{s} shows the behavior of the effective exponent $s_{\rm eff}=\frac{d\ln \gamma}{d\ln n}$. For the power law dependence of $\gamma$ the formal kinetic equation is
\begin{equation}\label{kin-2}
    \frac{dn}{dt}=-\gamma_0 t^{-\alpha} n^{2+s},
\end{equation}
having the solution
\begin{equation}\label{kin-2-sol}
    n(t)=\frac{n_0}{\left(1+\frac{1+s}{1-\alpha}\gamma_0 n_0^{1+s}t^{1-\alpha}\right)^{1/(1+s)}},
\end{equation}
with the asymptotic behavior for $t\rightarrow\infty$
\begin{equation}
n(t)\propto t^{-p}, \hskip10pt p=\frac{1-\alpha}{s+1}.
\label{p}
\end{equation}
We incorporate in \eq{kin-2} the possible effect of the dispersive transport leading to the power law time dependence of the recombination rate with $0\le \alpha < 1$.\cite{Pivrikas:242}

As it was already noted, the deviation of the recombination kinetics from the simple bimolecular decay $\propto 1/t$ is a possible argument in favor of the two-dimensional recombination, yet the dispersive transport with $\alpha > 0$ may imitate the same behavior. As we see, the effective exponent depends on the actual range of $n$. Experimental kinetics frequently may be characterized by $s_{\rm eff}= 0.5 - 1.5$\cite{Shuttle:93311,Shuttle:113201,Deibel:163303} and \fig{s} shows that the parallel disorder is more favorable for the development of $s > 0$ in the relevant density range.

\begin{figure}[tbp]
\centering
\includegraphics[width=3in]{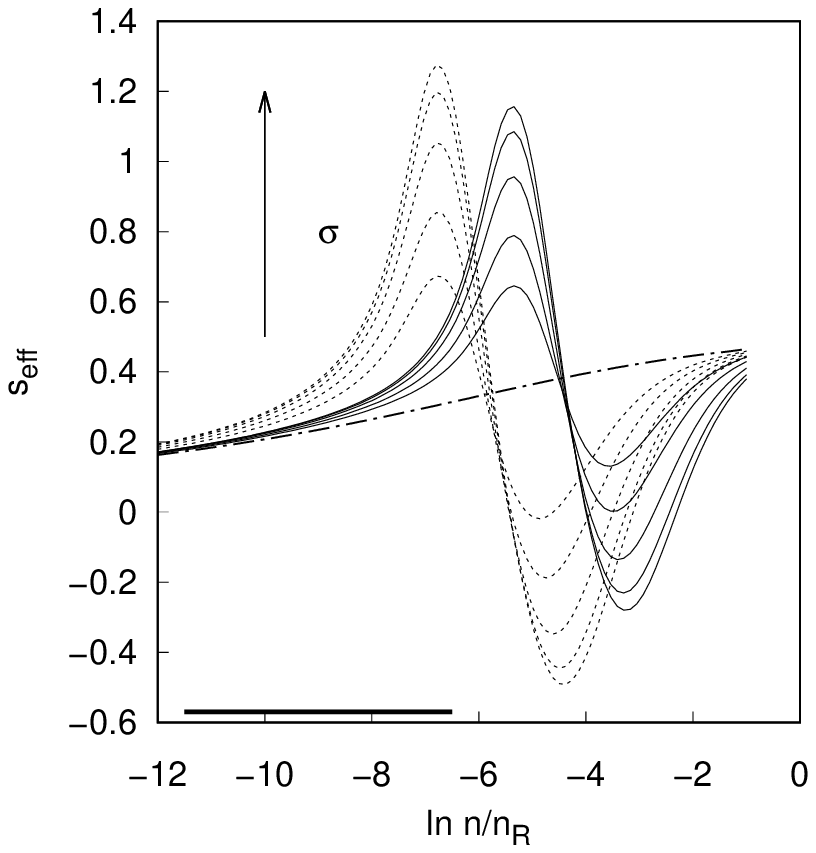}
\caption{Effective exponent $s_{\rm eff}$ for the parallel disorder and Gaussian  correlation function $C(r)=\exp\left(-r^2/2l^2\right)$ for $\sigma$ equals to 0.05 eV, 0.07 eV, 0.1 eV, 0.13 eV, and 0.15 eV, from the bottom curve upwards. Solid lines show $s_{\rm eff}$ for $l=5$ nm and dashed lines for $l=10$ nm, correspondingly. }
\label{n2}
\end{figure}

Typical charge density for the recombination experiments is $10^{15}-10^{17}$ cm$^{-3}$,\cite{Juska:13303,Sliauzys:224,Pivrikas:677}  and, assuming the typical space between lamellas as $\simeq 1.5-2$ nm,\cite{Sirringhaus:685} the two-dimensional density is $n\approx 10^8-10^{10}$ cm$^{-2}$. The limiting density $n_R$ is equal to $\simeq 10^{13}$ cm$^{-2}$. Hence, the typical ratio $n/n_R$ falls in the range $10^{-5}-10^{-3}$ (indicated in all figures) and in this range the two-dimensional recombination could generate a vast diversity of possible recombination kinetics. It is commonly believed that for the two-dimensional recombination without disorder effects $s=0.5$,\cite{Greenham:245301,Nenashev:213304} though the gradual transition to the logarithmic dependence (\ref{low-n}) for the relevant experimental density range should give much lower $s$. Disorder effects provide an additional complication. We see that $s$ could be even negative, thus providing the more faster decay in comparison to the usual bimolecular recombination. Development of the negative $s$ becomes easier for faster decaying correlation functions (see \fig{n2}). In combination with the dispersive transport negative $s$ can mimic the ordinary bimolecular recombination kinetics $n(t)\propto 1/t$.

\begin{figure}[tbp]
\centering
\includegraphics[width=3in]{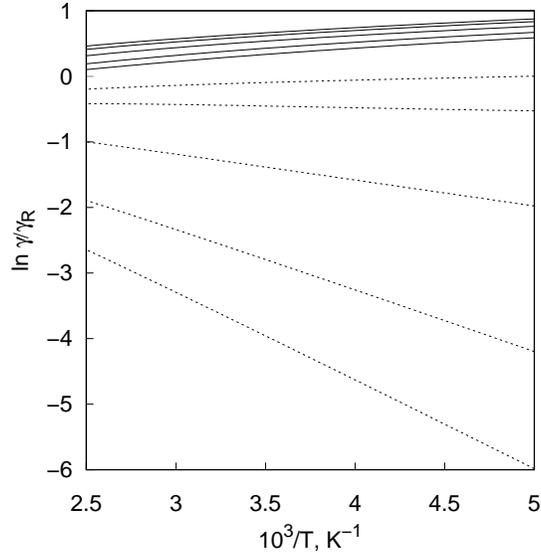}
\caption{Temperature dependence of $\gamma$ for the exponential correlation function with the correlation length $l=5$ nm, $\ln n/n_R=-7$, and $\sigma$ equals to 0.05 eV, 0.07 eV, 0.1 eV, 0.13 eV, and 0.15 eV. Solid lines show the dependences for the parallel disorder with the minus sign in $U_{\rm eff}(r)$ in \eq{gamma0-1} (the ratio $\gamma/\gamma_R$ weakly grows with $\sigma$), while the dashed lines show the corresponding dependences for the anti-parallel disorder with the plus sign in $U_{\rm eff}(r)$ (the ratio $\gamma/\gamma_R$ decays with $\sigma$). }
\label{T}
\end{figure}

\fig{T} shows the temperature dependence of the ratio $\gamma/\gamma_R$ for both types of disorder. As it was already noted, the temperature dependence for the case of the parallel disorder is very weak and the temperature dependence of $\gamma$ is mostly governed by the diffusivity dependence on $T$, while for the anti-parallel disorder this additional dependence is strong.

In the annealed polythiophene-based solar cells the reduction factor $\zeta$ decays with temperature, the behavior hardly explainable by the known models.\cite{Juska:1167,Deibel:163303}
OPV materials discussed in these papers are bulk heterojunction materials, yet there is a strong indication that the recombination takes place in lamellas and its characteristics are not determined by the interfaces between donor and acceptor components. In particular, many features of the recombination process in materials containing blends of regio-regular poly(3-hexylthiophene) (P3HT) as hole-transporting component are common with the carrier recombination in TiO$_2$/P3HT structures without mesoscopic segregation.\cite{Juska:13303} General behavior of $\zeta(T)$ can be naturally explained by the two-dimensional recombination in the lamellar structure (obviously, the annealing is favorable for the formation of lamellas). If we rewrite $\gamma_{\rm L}$ using the diffusivity, then
\begin{equation}\label{zeta}
\zeta = \frac{D_{2} B(T)}{D_{3} R_{\rm Ons}}\propto TB(T)\exp\left[-\frac{1}{2}\left(\frac{\sigma_{2}}{kT}\right)^2+
\frac{1}{3}\left(\frac{\sigma_{3}}{kT}\right)^2\right],
\end{equation}
where $D_2$ is the two-dimensional diffusivity in lamella, $D_3$ is the macroscopic three-dimensional transport diffusivity, $B(T)$ is the recombination related factor, and the leading asymptotics for the temperature dependence of the diffusivity for the Gaussian density of states in the $d$-dimensional space is used\cite{Deem:911,Novikov:275}
\begin{equation}\label{D_d}
D_d\simeq D_0\exp\left[-\frac{1}{d}\left(\frac{\sigma}{kT}\right)^2\right].
\end{equation}
The crucial moment in this estimation is that the two-dimensional $\gamma$ is controlled by the lamellar disorder $\sigma_{2}$ while the macroscopic three-dimensional transport is determined by the different $\sigma_{3}$ and due to the local order in lamellas it is natural to expect that $\sigma_{2} < \sigma_{3}$. Then
\begin{equation}\label{zeta-T}
\frac{d\ln\zeta}{d \ln T} \simeq 1+\frac{d\ln B}{d\ln T}+\left(\frac{\sigma_{2}}{kT}\right)^2-\frac{2}{3}\left(\frac{\sigma_{3}}{kT}\right)^2
\end{equation}
Experiment gives $\frac{d\ln\zeta}{d \ln T}\approx -5.7$.\cite{Juska:1167,Deibel:163303} According to \eq{zeta-T}, if we use the experimental value for $\sigma_3\approx 0.066$  eV,\cite{Mauer:2085} recalculated from the GDM value\cite{Bassler:15} according to the more proper estimation (\ref{D_d}), then for $T=250K$ and $B\simeq {\rm const}$ the proper value of the derivative can be obtained for $\sigma_2\approx 0.035$ eV. This estimation looks plausible. Again, the parallel disorder is more favorable for the negative value of the derivative (\ref{zeta-T}) due to moderate negative values of $\frac{d\ln B}{d\ln T}$, while for the anti-parallel disorder $\frac{d\ln B}{d\ln T}$ is large and positive (see \fig{2d-exp}, the increase of $\sigma$ is roughly equivalent to the decrease of $T$). At the same time, development of the significant conformational contribution to the total energetic disorder seems possible for the regio-regular polythiophene where electrostatic contribution should be small (more detailed discussion may be found elsewhere\cite{Novikov:18854}). Hence, the two-dimensional recombination with the parallel energetic disorder gives the plausible explanation for $d\zeta/dT <0$  in the annealed polythiophene-based devices.

\section{Conclusions}
In conclusion, we presented the theory of the two-dimensional recombination in amorphous organic semiconductors taking into account the spatial correlation of the random energy landscape. Correlated disorder strongly affects the recombination kinetics and naturally explains the temperature dependence of the recombination reduction factor $\zeta$.
Our results hint that the effective exponent $s_{\rm eff} \simeq 0.5 -1.0$ and the corresponding slow decay kinetics for the non-dispersive carrier transport and relevant carrier density are hardly achievable without support from the correlated disorder. If the disorder is absent, then even the classical exponent $s=0.5$\cite{Juska:13303} could be achieved only in a very close vicinity of the limiting density $n_R=1/\pi R^2$, inaccessible in real experiments.

Alternative and more traditional explanation of the large exponent $s \simeq 0.5 - 4$ is the effect of the mesoscopic phase separation in OPV materials.\cite{Credgington:1465,Nenashev:33301,Nenashev:104207,Albes:20974} Yet in some situations the bulk heterojunction material (for example, the blend of  poly[2-methoxy-5-(3,7-dimethyloctyloxy)-
phenylene vinylene] (MDMO-PPV) mixed with  1-(3-methoxycarbonyl)propyl-
1-phenyl-[6,6]-methanofullerene (PCMB)) demonstrates the perfect Langevin recombination,\cite{Pivrikas:677} though having the similar heterogeneous structure to those of the blend of  regio-regular P3HT and PCMB,\cite{Douheret:32107} thus again hinting that the specific two-dimensional structure of P3HT cells and not the mesoscopic segregation of the donor and acceptor components is responsible for the non-Langevin recombination.

Our results show that the two-dimensional recombination in the spatially correlated random landscape could provide the comparable rich behavior depending on the disorder parameters and actual carrier density range. Extremely interesting is the possibility to realize the regime with $s < 0$; this regime may be already observed in neat P3HT.\cite{Gorenflot:144502} The authors described the observed recombination kinetics as the mixture of the contributions from the second- and first-order processes; it could be equally well described by $s < 0$ and small $|s|$. For the models derived previously by Nenashev \textit{et al}.\cite{Nenashev:33301,Nenashev:104207} such behavior could be realized only for the unreasonably large carrier density. Moreover, it is not very clear how accurately the model of one-dimensional domains could capture the diffusive transport and recombination in real bulk heterojunction materials.

Probably the most important conclusion from the considered model is that the two-dimensional recombination in the random environment is capable to produce very different effective exponents $s_{\rm eff}$, thus the frequently used reason for the invocation of the conception of the two-dimensional recombination for the explanation of the slow-decaying kinetics with $s_{\rm eff} \simeq 0.5$ or greater should be modified. We see that even observation of $s_{\rm eff}\approx 0$ or $s_{\rm eff} < 0$ does not exclude the possibility that the actual recombination process takes place in some two-dimensional formation, though, of course, the ordinary Langevin recombination with some modifications is a more natural mechanism.

We should note also that the method considered in this paper could be applied to problems related to the bimolecular reaction in the 2D geometry in the presence of the disorder, as various as charge recombination in the Langmuir monolayers,\cite{Wittek:870,Vuorinen:5383} cell recombination,\cite{Zhu:23} enzymes reaction at the cell membranes,\cite{Berry:1891} and catalytic reactions at the amorphous surfaces of solid catalysts.\cite{Ertl:3524}

\section*{Acknowledgements}
Financial support from the Ministry of Science and Higher Education of the Russian Federation (A.N. Frumkin Institute) and Program of Basic Research of the National Research University Higher School of Economics is gratefully acknowledged.

\bibliography{recombination-2d} 

\end{document}